\documentclass[10pt]{article}
\usepackage{geometry}
\usepackage{cite}
\usepackage{amssymb,amsmath}
\usepackage{wrapfig}
\usepackage{graphicx}
\usepackage[small]{caption}
\usepackage[titletoc,title]{appendix}
\usepackage{authblk}

\begin{document} 
\title{Effect of Particle Number Conservation on the Berry Phase Resulting from Transport of a Bound Quasiparticle around a Superfluid Vortex}
\author[1]{Yiruo Lin \thanks{yiruolin@illinois.edu}}
\author[1,2]{Anthony J. Leggett \thanks{aleggett@illinois.edu}}
\affil[1]{\textit{Department of Physics, University of Illinois at Urbana-Champaign}}
\affil[2]{\textit{Shanghai Center for Complex Physics, Shanghai Jiaotong University}}

\date{}
\maketitle

\begin{abstract}
Motivated by understanding Majorana zero modes in topological superfluids in particle-number conserving framework beyond the present framework, we study the effect of particle number conservation on the Berry phase resulting from transport of a bound quasiparticle around a superfluid vortex. We find that particle-number non-conserving calculations based on Bogoliubov-de Gennes (BdG) equations are unable to capture the correct physics when the quasiparticle is within the penetration depth of the vortex core where the superfluid velocity is non-zero. Particle number conservation is crucial for deriving the correct Berry phase in this context, and the Berry phase takes non-universal values depending on the system parameters and the external trap imposed to bind the quasiparticle.  Of particular relevance to Majorana physics are the findings that superfluid condensate affects the part of the Berry phase not accounted for in the standard BdG framework, and that the superfluid many-body ground state of odd number of fermions involves superfluid condensate deformation due to the presence of the bound quasiparticle - an effect which is beyond the description of the BdG equations.
\end{abstract}

\section{Introduction}

In this paper, we address the following basic question: What is the Berry phase of transporting a bound quasiparticle around a superfluid vortex?  Surprisingly, this basic question has not been systematically considered in the literature to the best of our knowledge. It is of particular interest to us to understand possible many-body effects, in particular the role played by the superfluid condensate in determining the Berry phase, thereby examining the validity of Bogoliubov-de Gennes (BdG) equations \cite{BdG} for calculating the Berry phase. \\

This work is motivated by our recent study \cite{Lin_Leggett_1} on effect of particle number conservation on the Berry phase in braiding Majorana zero modes (for a review on Majorana zero modes, see e.g., \cite{Alicea_rev}). The currently established framework for studying Majorana zero modes in superfluids relies on the BdG equations which break particle number conservation. On the other hand, relevant physical quantities such as Berry phase are determined by many-body quantum states that conserve fermion number. Furthermore, the effect of the superfluid condensate  may not be neglected when it has interesting structure (see for example, the NMR of superfluid Helium-3\cite{Leggett_1} and of surface of B phase superfluid Helium-3 hosting Majorana zero modes \cite{Silaev}). The superfluid condensate needs to have nontrivial topology for the existence of Majorana zero modes. Therefore, it is necessary to examine the validity of the present framework by studying effect of particle number conservation and condensate contribution to Majorana physics. We are particularly interested in understanding braiding statistics of Majorana zero modes, for which a basic physical quantity involved is Berry phase of transporting a bound quasiparticle around a superfluid vortex, the subject of in this work. To simplify physics, we restrict ourselves to simple s-wave superfluids and suppose that the bound quasiparticle is localized by an external Zeeman potential. Furthermore, we adopt an annular geometry for confining the superfluid to make the problem effectively one-dimensional. The vortex is then simulated by quantized winding number of Cooper pairs in the superfluid around the annulus. \\

The paper is organized as follows. In section \ref{list}, we define the problem and state the main results. In section \ref{BdG}, we calculate the Berry phase in the framework based on the BdG equations in particle-number non-conserving approximation. In section \ref{exact}, we relate the Berry phase to the system angular momentum. In section \ref{sum rules and continuity condition}, we discuss the effect of particle number conservation on the Berry phase in terms of the system angular momentum in linear response theory. In particular, we propose a many-body ground state ansatz beyond the BdG description in order to satisfy the continuity condition which is necessary for compatibility with the f-sum rule. In section \ref{lower bound}, we estimate lower bound on the Berry phase at non-zero superfluid velocities to show the contribution from the superfluid condensate when the superfluid is moving. In section \ref{square}, we carry out explicit calculations of the Berry phase for a particular form of the Zeeman potential and obtain analytical expression at general superfluid velocities. Finally in section \ref{summary}, we summarize our results and draw our conclusions on the effect of particle number conservation on the Berry phase. \\

\section{Annulus model and main results} \label{list}

\begin{figure}[h!]
\begin{center}
\includegraphics[width=0.35\textwidth]{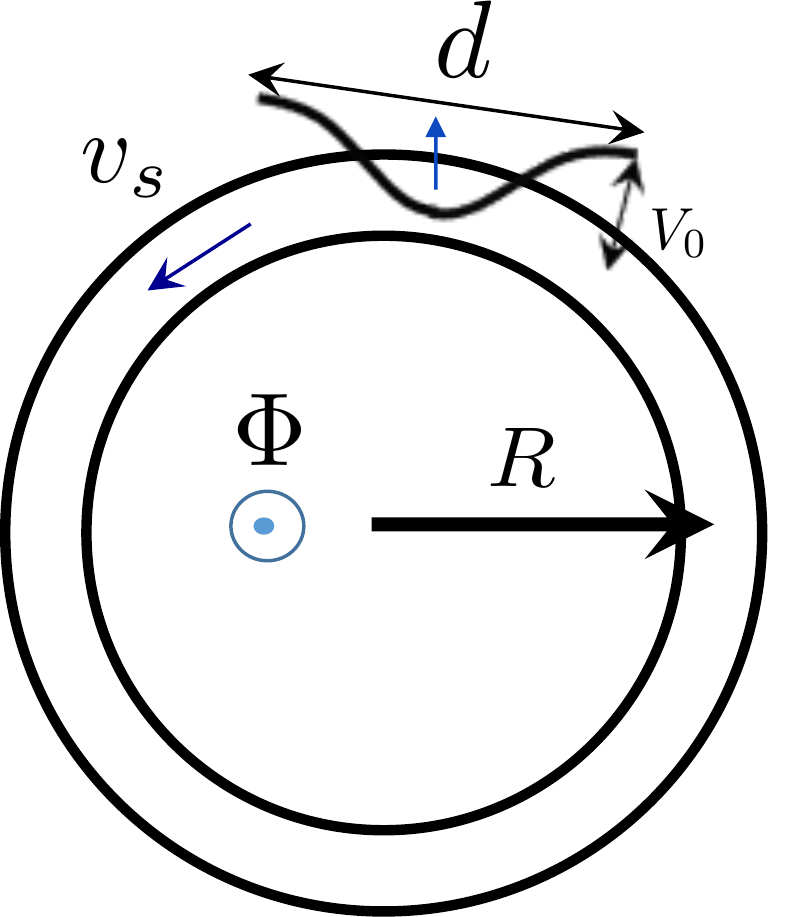}
\end{center}
\caption{A bound quasiparticle with spin pointed along the direction of Zeeman field is formed in a s-wave superfluid with odd number of particles confined in an annular geometry. $d$ is spatial extension of Zeeman field, $V_0$ is its characteristic strength and the superfluid velocity is given by $v_s=(n+2\Phi)\hbar/2mR$, where $n$ is the condensate winding number, $\Phi$ is the magnetic flux through the annulus in units of $h/|e|$, $R$ is radius of the annulus. }
\label{annulus}
\end{figure}

The system we consider is described in ref.[6]: a BCS s-wave superfluid of fermions (which we now equip with charge e to permit interaction with an external magnetic flux) with total fermion number 2N+1, confined in the annular geometry shown in figure \ref{annulus}. The system doesn't exchange particles with its environment, hence the total fermion number is fixed. Due to the odd particle-number parity, there is an unpaired fermion in the ground state. In the framework of the BdG equations, the system ground state can be approximated by the first excited eigenstate obtained in the particle-number non-conserving approximation followed by a projection onto fixed (odd) particle-number sector. In this physical picture, there is a quasiparticle in its lowest excited state in the superfluid condensate. We shall adopt such a picture and consider the effect of particle number conservation on it. To bound the quasiparticle, a weak Zeeman field is imposed on a segment of annulus whose characteristic size along the annulus is much larger than the superfluid coherent length. The Zeeman field has no variation in radial direction so we can treat the bound quasiparticle in an effective one dimensional potential. The Zeeman field is weak enough so that corresponding even particle-number parity ground state can be approximated by the homogeneous BCS s-wave ground state. The justification of the aforementioned approximations can be found in \cite{Lin_Leggett}. To study the effect of a non-zero superfluid velocity on the Berry phase, a magnetic flux is inserted so that the superfluid velocity can be tuned continuously by the flux while keeping the condensate winding number fixed. To realize adiabatic evolution, we need to move the Zeeman trap slowly enough so that the system stays in the ground state of the instantaneous Hamiltonian at any moment. This can be achieved as long as the timescale of transporting the Zeeman trap is much larger than other system timescales related to system excitations. In particular, for a neutral Fermi superfluid, the timescales associated with low-energy phonons increase with system size and scale as $R/c\sim R/v_F$ ($v_F$ is Fermi velocity). For such a system, the time scale for moving the Zeeman trap should be much larger than $R/v_F$. So the linear speed of dragging the trap decreases as system size. Nevertheless, we can always realize adiabatic evolution and observe the Berry phase. \\

Results: We found that the Berry phase at vanishing superfluid velocity is equal to Aharonov-Bohm (AB) phase. For a vortex of unit winding number, the Berry phase is equal to $\pi$, consistent with the standard result. However when the superfluid velocity is non-zero (which can occur when for instance, the quasiparticle is within the penetration depth from the vortex core in a superconductor), the Berry phase receives contributions from the superfluid condensate and becomes non-universal. Particle number conservation needs to be respected for calculating the Berry phase at non-zero superfluid velocities. In the annulus model, we found evidence for the modification necessary for obtaining the correct Berry phase of the system many-body ground state beyond the description given by the BdG equations. The modification involves entanglement between the quasiparticle wave function and the condensate wave function. Such a modification goes beyond the single-particle picture described in the BdG framework and implies the important role played by the many-body effect due to the superfluid condensate. The proposed modification can shed light on properties of Majorana zero modes beyond the description given by the BdG equations.

\section{Berry phase calculated in BdG formalism in particle-number non-conserving approximation}\label{BdG}

\begin{figure}[h!]
\begin{center}
\includegraphics[width=0.3\textwidth]{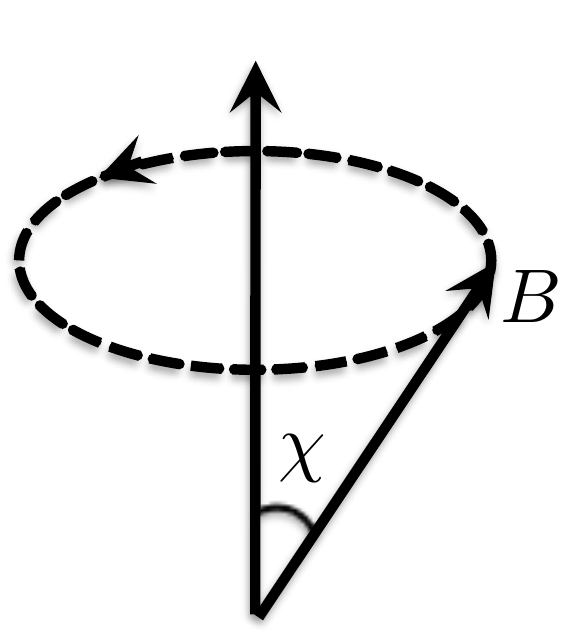}
\end{center}
\caption{A bound quasiparticle as an effective spin in a magnetic field. For the bound quasiparticle, effective $\chi=\pi/2$, and Berry phase is $\phi= 2n\pi\mathrm{cos}^2(\chi/2)=n\pi$. }
\label{fig_spin}
\end{figure}

In the standard mean-field approximation, particle number conservation is broken down to $Z_2$ symmetry. In the BdG framework, we may view the condensate as vacuum and regard the system in consideration as an effective single particle with pseudo-spin degree of freedom representing particle and hole component of the quasiparticle. Taking the analogy of the BdG equations for the quasiparticle to the Schrodinger equation for a spin-1/2 degree of freedom in a magnetic field illustrated in figure \ref{fig_spin}, we can map the kinetic energy of the particle and the hole to the z-component of the effective magnetic field and superfluid gap plays the role of the magnetic field in the x-y plane. Namely, the BdG equation for the quasiparticle 
\begin{eqnarray}
\left(\begin{array}{cc}H_{\uparrow} & \bigtriangleup(r) \\ \bigtriangleup^*(r) & -H_{\downarrow} \end{array}\right)\left(\begin{array}{c}u_{\uparrow}(r) \\v_{\downarrow}(r)\end{array}\right) =E\left(\begin{array}{c}u_{\uparrow}(r) \\v_{\downarrow}(r)\end{array}\right), \label{BdG_eq}
\end{eqnarray}
where $\bigtriangleup(r)=|\bigtriangleup|\mathrm{e}^{in\theta}$ with $\theta$ azimuthal coordinate along the annulus, can be identified with the Schrodinger equation of a spin in a rotating magnetic field
\begin{eqnarray}
\left(\begin{array}{cc}B_z & B(r) \\ B^*(r) & -B_z \end{array}\right)\left(\begin{array}{c}s_\uparrow \\ s_\downarrow\end{array}\right) =E\left(\begin{array}{c}s_\uparrow \\s_\downarrow\end{array}\right). \label{spin_Schrodinger}
\end{eqnarray} \\
It is worth noting that equation (\ref{BdG_eq}) is not the exact analog of equation (\ref{spin_Schrodinger}), since $H_{\uparrow}$ is not equal to $H_\downarrow$ (cf.equation (\ref{BdG_1/2_B}) and (\ref{BdG_0_transf})). However, the difference is proportional to the unit matrix in Nambu space and hence should not affect the argument. 

As the quasiparticle is moved around the annulus and returned to its starting point, the local superconducting gap phase seen by the quasiparticle is wound by $2n\pi$ where $n$ is the winding number of the vortex and so the effective magnetic field is rotated about the $z$ axis by $2n\pi$. For a bound state, the particle and the hole component have the same weight (this can be seen as follows: the particle component gets reflected back at the trap edge and becomes the hole which subsequently gets reflected back as the particle; since the quasiparticle is trapped, the particle and the hole must have the same weight: if we "observe" the quasiparticle, there is equal probability of finding it in the particle and in the hole state), so the effective spin lies in the x-y plane. From the well-known result of the Berry phase of the spin-1/2 particle in the magnetic field, we immediately get the Berry phase to be $n\pi$ \cite{Leggett_Lin}. This result appears to be insensitive to whether there is magnetic flux through the annulus. We will now turn to an alternative approach which gives a different answer.

As will be shown in the next section, the Berry phase can be found from the difference in the total angular momentum between $2N+1$-particle and $2N$-particle ground states. Making use of continuity condition, we know that in a one-dimensional system, the current is uniform throughout the annulus for any energy eigenstate. Writing the $2N+1$-particle ground state as the BdG quasiparticle operator $\alpha^\dagger$ acting on the $2N$-particle ground state, we can write the Berry phase in terms of commutator between current density operator and $\alpha^\dagger$ at any arbitrary position $\theta'$
\begin{eqnarray}
\phi/2\pi&=&L(\langle\alpha \tilde{J}(\theta')\alpha^\dagger\rangle-\langle \tilde{J}(\theta')\rangle)-\Phi \nonumber \\
&=&L(\langle\alpha[\tilde{J}(\theta'),\alpha^\dagger]\rangle)-\Phi, \label{commut}
\end{eqnarray}
where $L$ is annulus circumference, $\Phi$ is the magnetic flux through the annulus (see figure \ref{annulus}) which appears in the above expression due to converting the angular momentum density $J(\theta')$ to current $\tilde{J}(\theta')$ at $\theta'$ by $\tilde{J}(\theta')=J(\theta')+\Phi\rho(\theta')/L$ (throughout the paper, "angular momentum" refers to canonical angular momentum and "current" refers to kinetic current which is proportional to particle velocity). In the particle number non-conserving form, $\alpha^\dagger=\int d\theta u(\theta)\Psi^\dagger(\theta)+v(\theta)\Psi(\theta)$ where $u(\theta)$ and $v(\theta)$ are the particle and the hole wave functions localized at the Zeeman trap around $\theta_0$ ($\theta$ is azimuthal angle along the annulus). We can always choose $\theta'$ to be sufficiently far away from $\theta_0$ such that $[\tilde{J}(\theta'),\alpha^\dagger]=0$. So the Berry phase is just given by AB phase at any magnetic flux. This result is in conflict with the previous result obtained using the spin analogy in which the Berry phase is found to be independent of the magnetic flux. In deriving equation (\ref{commut}), the key step is to make use of the continuity condition to replace the total angular momentum by the local current. Although the continuity condition is satisfied in the BdG formalism for any energy eigenstate or at thermal equilibrium, its validity in those situations is justified only under the particle-number non-conserving approximation. On the other hand, we know that the system in consideration has fixed particle number, which is necessary for a physically meaningful quantum phase associated with the adiabatic evolution. So the Berry phase $-\Phi$ obtained from (\ref{commut}) doesn't correspond to a quantum state with fixed particle number and may not correspond to a physical result. The unphysical result will be modified already at a naive level of restoring particle number conservation. Once we add a Cooper pair creation operator associated with the hole part of the BdG quasiparticle creation operator (cf. equation (\ref{alpha_BdG})), the commutator  $[\tilde{J}(\theta'),\alpha^\dagger]$ becomes finite throughout the annulus as the Cooper pair wave function is spread out along the annulus. \\

\section{Berry phase in terms of angular momentum} \label{exact}

In this section, we develop a formalism in which the Berry phase is related to the system angular momentum. We prove that the exact value of the Berry phase can be obtained at vanishing superfluid velocity. In linear response approximation, we evaluate the Berry phase at non-zero superfluid velocities in terms that depend only on the ground state wave function and the system energy spectrum at zero superfluid velocity. The formula for non-zero superfluid velocities derived in this section will be studied in the following two sections. \\

In an effective one-dimensional system, the Berry phase is related to the total angular momentum of the system. This is because we can write the system many-body ground state in the form $\Psi(\{\theta_i-\theta_0\})$, where only azimuthal coordinate is explicitly considered, $\theta_i$ is the coordinate of particle $i$ which runs from 1 to $2N+1$ and $\theta_0$ parameterizes the position of the Zeeman trap. The Berry phase can then be shown to be equal to the total angular momentum of the system as follows
\begin{eqnarray}
\phi&=&-\mathrm{Im}\{\int_0^{2\pi}\langle\Psi|\frac{\partial\Psi}{\partial\theta_0}\rangle\} \nonumber \\
&=&\mathrm{Im}\{\sum^{2N+1}_{i=1}\int_0^{2\pi}\langle\Psi|\frac{\partial\Psi}{\partial\theta_i}\rangle\} \nonumber \\
&=& 2\pi\sum_{i=1}^{2N+1}L_i, \label{Berry}
\end{eqnarray}
where $L_i$ denotes the expectation value of the angular momentum of particle $i$. \\

It is worth noting that the total quantum phase of interest accumulated in the adiabatic evolution is the sum of the explicit phase (which is usually called monodromy phase) of the instantaneous ground state and the Berry phase. In the above derivation, the monodromy phase is zero since the instantaneous ground state can be chosen to take the form $\Psi(\{\theta_i-\theta_0\})$ which explicitly returns to its initial state after the adiabatic evolution when $\theta_0$ winds by $2\pi$. This is due to one-dimensional nature of the problem: changing $\theta_0$ by $2\pi$ is completely equivalent to changing $\theta_i$ by $2\pi$, the effect of which on the ground state must be to return it to its initial state by single-valued condition. Therefore, for our system, the total quantum phase is just equal to the Berry phase. Note that this is not true in general situations where we need to calculate both the monodromy phase and the Berry phase. For example, in the case of Majorana zero modes in vortices of p+ip superfluids, the relative monodromy phase of interchanging two Majorana zero modes is non-zero for degenerate ground states \cite{Ivanov}. \\

It turns out that at integer or half-integer magnetic fluxes (measured in unit of $h/|e|$), we can find the angular momentum exactly based on general arguments from gauge invariance and time reversal symmetry. Away from the integer or half-integer fluxes, the angular momentum can't be found exactly and we need to resort to first-order perturbation theory.

\subsection{Integer or Half-integer Flux} \label{Integer Flux}

We write down the system Hamiltonian in the following form
\begin{eqnarray}
H&=&\sum_{j=1}^{2N+1}(-i\frac{\partial}{\partial\theta_j}+\Phi)^2+V_{\mathrm{int}}+H_z, \label{H}
\end{eqnarray}
where $V_{\mathrm{int}}$ refers to particle-particle interaction, $\Phi$ is the external magnetic flux in units of $h/|e|$ and $H_z$ is the Zeeman term. Making a gauge transformation to the $2N+1$-particle wave function $\Psi$
\begin{eqnarray}
\tilde{\Psi}=\mathrm{exp}(i\sum^{2N+1}_{j=1}\Phi\theta_i)\Psi \label{Gauge_Psi}
\end{eqnarray}
and substituting into the Schrodinger equation, we get 
\begin{eqnarray}
\tilde{H}\tilde{\Psi}=E\tilde{\Psi}, \label{Gauge_Schrodinger}
\end{eqnarray}
where $\tilde{H}=-\sum_{j=1}^{2N+1}\partial^2/\partial\theta_j^2+V_{\mathrm{int}}+H_z $. The transformed many-body function satisfies the boundary condition
\begin{eqnarray}
\tilde{\Psi}(\theta_i+2\pi,...)=\mathrm{exp}(i2\pi\Phi)\tilde{\Psi}(\theta_i,...). \label{BC_Gauge}
\end{eqnarray}
When $2\Phi=n$, the boundary condition (\ref{BC_Gauge}) after the gauge transformation is invariant under time reversal. Since the transformed Hamiltonian $\tilde{H}$ also has time reversal symmetry (note that the Zeeman term is unchanged since the time reversal considered here is with respect to orbital degrees of freedom only), the ground state wave function $\tilde{\Psi}$  must be real if there is no degeneracy. So its angular momentum is zero. Thus, the angular momentum of the original wave function is
\begin{eqnarray}
L=-\sum_{i=1}^{2N+1}\Phi=-(2N+1)\Phi. \label{L}
\end{eqnarray}

Similarly for even number of particles, the angular momentum is $-2N\Phi$. Note that the above result (\ref{L}) is very general and exact. Furthermore, it doesn't depend on details of the system such as whether it's superconducting or not. It is also different from general theorem of Byers and Yang \cite{BY} in that it is a stronger statement on non-degenerate eigenstates at integer fluxes of $h/2|e|$, instead of $h/|e|$. It is rather interesting to see that the fluxes quantized at integer values of $h/2|e|$ are special in that the Berry phase of transporting any local potential (in the case of current interest, it is the Zeeman field) becomes just the sum of AB phase of each individual particle moving around the annulus. Usually, $h/2|e|$ is related to Cooper pairing. However, here it enters in quite general situations. \\

Equation (\ref{L}) can be applied to our toy model when the superfluid velocity vanishes. Since at zero superfluid velocity,  we have
\begin{eqnarray}
l_0+2\Phi=0, \label{zero_v}
\end{eqnarray}
where $l_0$ is the superfluid winding number. For vortex winding number $l_0=1$, $\Phi=-1/2$ which is at half-integer value of $h/|e|$. Furthermore, we know the ground state is unique for our system with a single bound quasiparticle in potential well (the energy from spin degree of freedom is split due to Zeeman field). According to (\ref{L}) and (\ref{Berry}), the Berry phase is equal to $\pi$. Thus the calculations in Section \ref{BdG} in the BdG approximations are correct at zero superfluid velocity. We shall see in the following that they cease to be correct when the superfluid is moving. \\

\subsection{Linear Response Theory - away from Integer or Half-integer Fluxes}

Away from the integer or half-integer fluxes, the boundary condition (\ref{BC_Gauge}) is no longer invariant under time reversal and the above argument ceases to be valid. In order to proceed, we can regard the deviation of magnetic flux from integer or half-integer values as perturbation and apply first-order perturbation theory to the problem. This is valid if the annulus is large enough so that $O(1)$ flux change is a small perturbation to the system compared to total energy of the system.\\

To obtain the Berry phase that characterizes the quasiparticle statistics, we should compare the difference in the Berry phase for system with $2N+1$ and $2N$ particles. Taking this into account and writing the Berry phase in terms of deviation from the value from integer or half-integer magnetic flux, we get the following expression for the Berry phase at magnetic flux $\Phi=-1/2+\delta\Phi$ (from now on, we fix the superfluid winding number to be 1)
\begin{eqnarray}
\phi/2\pi&=&(\langle J_{2N+1}\rangle_{\Phi=-\frac{1}{2}+\delta\Phi}-\langle J_{2N+1}\rangle_{\Phi=-\frac{1}{2}})-(\langle J_{2N}\rangle_{\Phi=-\frac{1}{2}+\delta\Phi}-\langle J_{2N}\rangle_{\Phi=-\frac{1}{2}}) +\frac{1}{2}, \label{phi_J}
\end{eqnarray}
where $J\equiv\sum_j-i\partial/\partial\theta_j$. Note that we use notation $J$ here to emphasize relation between the Berry phase and current response to transverse vector field acting on the superfluid due to the magnetic flux $\Phi$ threading the annulus. $J$ is just the total angular momentum of the system (it is the same quantity as the $L$ appearing in equation (\ref{L})). \\

Now applying standard first-order perturbation theory, the angular momentum $J$ difference at flux $\Phi$ and at flux $-1/2$ is
\begin{eqnarray}
\langle\tilde{0}|J|\tilde{0}\rangle-\langle0|J|0\rangle&=&(\sum_na_n\langle0|J|n\rangle+c.c.) \nonumber \\
&=&2\sum_n\frac{|\langle0|J|n\rangle|^2}{E_0-E_n}\delta\Phi, \label{LR}
\end{eqnarray}
where $|\tilde{0}\rangle$ refers to the ground state at flux $\Phi=-\frac{1}{2}+\delta \Phi$, $|0\rangle$ and $|n\rangle$ refer to ground state and excited eigenstates at flux $\Phi=-\frac{1}{2}$ with energies $E_0$ and $E_n$, respectively, $|\tilde{0}\rangle=|0\rangle+\sum_na_n|n\rangle$. We omit the subscript of the angular momentum $J$, which will be added for discussing cases with $2N$ and $2N+1$ particles separately (Section \ref{sum rules}). The RHS of the above equation gives the modification to the Berry phase away from the integer or half-integer fluxes, corresponding to non-zero superfluid velocities.  \\

\section{Sum rules and continuity condition} \label{sum rules and continuity condition}

\subsection{Sum rules} \label{sum rules}

Let's now evaluate the sum in equation (\ref{LR}). Let's first discuss the $2N$-particle ground state.  If we assume the $2N$-particle ground state at flux $\Phi=-\frac{1}{2}+\delta\Phi$ has rotation symmetry, which is true to the first order of the Zeeman field strength, then the matrix element $\langle0|J|n\rangle$ vanishes identically and hence the correction due to $\delta\Phi$ in equation (\ref{LR}). So we have $\langle J_{2N}\rangle_{\Phi=-\frac{1}{2}+\delta\Phi}-\langle J_{2N}\rangle_{\Phi=-\frac{1}{2}}=0$ (recall that $J$ is the canonical angular momentum). This result is simply the Meissner effect which implies that the current-current correlation here is the transverse one and the superfluid condensate doesn't contribute to it. This is intuitively reasonable since the current is responding to magnetic vector potential as in the usual Meissner effect, which is analogous to the behavior of a superfluid in a rotating container\cite{AJL}. \\

What about the ground state with $2N+1$ particles? In this case, the ground state no longer has rotation symmetry, so the matrix element in equation (\ref{LR}) is finite. The sum rule of the current-current correlation here is rather tricky since we are considering ground state with odd total number of particles and it's unclear what accounts for normal fluid which is responsible for transverse current-current correlation. It is tempting to rewrite the sum in equation (\ref{LR}) as
\begin{eqnarray}
2\sum_n\frac{|\langle0|J_{2N+1}|n\rangle|^2}{E_0-E_n}&=&-i\frac{\langle0|[X, H]+\frac{i}{2}|n\rangle \langle n|J_{2N+1}|0\rangle}{E_0-E_n}-i\frac{\langle0|J_{2N+1}|n\rangle\langle n|[X, H]+\frac{i}{2}|0\rangle}{E_0-E_n}, \nonumber \\
\label{LR_1}
\end{eqnarray}
where $H$ is the Hamiltonian at flux $\Phi=-1/2$, $X=\sum_{i=1}^{2N+1}\theta_i$. The constant $i/2$ in the matrix element in equation (\ref{LR_1}) comes from the fact that the kinetic term for each particle $i$ in $H$ takes the form $(-i\frac{\partial}{\partial\theta_i}-\frac{1}{2})^2$. The first sum in equation (\ref{LR_1}) can be rewritten as
\begin{eqnarray}e
-i\frac{\langle0|[X, H]+\frac{i}{2}|n\rangle \langle n|J_{2N+1}|0\rangle}{E_0-E_n}=i\langle0|X|n\rangle\langle n|J_{2N+1}|0\rangle.  \label{LR_1-1}
\end{eqnarray} 
Similarly, the second sum in equation (\ref{LR_1}) can be written as
\begin{eqnarray}
-i\frac{\langle 0|J_{2N+1}|n\rangle\langle n|[X, H]+\frac{i}{2}|0\rangle}{E_0-E_n}=-i\langle 0|J_{2N+1}|n\rangle \langle n|X|0\rangle. \label{LR_1-2}
\end{eqnarray}
Now, adding equation (\ref{LR_1-1}) and (\ref{LR_1-2}), together with a term $i\langle0|X|0\rangle\langle0|J_{2N+1}|0\rangle-i\langle0|J_{2N+1}|0\rangle\langle0|X|0\rangle$ (which is zero), we get
\begin{eqnarray}
2\sum_n\frac{|\langle0|J_{2N+1}|n\rangle|^2}{E_0-E_n}&=&i\langle0|[X,J_{2N+1}]|0\rangle \nonumber \\
&=&-(2N+1). \label{LR_f}
\end{eqnarray} 
So the Berry phase is, combing equation (\ref{phi_J}) with equation (\ref{LR_f}) and (\ref{LR})
\begin{eqnarray}
\phi/2\pi=-(2N+1)\delta \Phi+\frac{1}{2}. \label{Berry_f}
\end{eqnarray}

However, the result (\ref{Berry_f}) is suspect, since if were to apply  the same derivation to the $2N$-particle ground state, we would arrive at Berry phase change away from $\Phi=-1/2$ proportional to $2N$ which conflicts with the earlier argument based on properties of the many-body ground state.  This is clearly the analog, in our geometry, of the well-known failure, in a bulk system, of the longitudinal and transverse current-current correlations to coincide in the static $q\rightarrow0$ limit \cite{Baym}. Mathematically, we can't simply take the commutation relation in (\ref{LR_f}) to be that of $[x,p]$ as here $X$ is compact so that it is defined modulo $2(2N+1)\pi$. Nevertheless, we may regard the (magnitude of ) right hand side of equation (\ref{LR_f}) as an upper bound of (the magnitude of) the sum on the left hand side. In the following, we'll make various attempts to estimate the sum. We'll see that the standard BdG framework results in violation of the upper bound and it's necessary to modify the corresponding many-body wave functions beyond the BdG construction to restore particle number conservation and to comply with the sum rule given by equation (\ref{LR_f}). \\

\subsection{Continuity condition} \label{continuity condition}
 
So far we have not considered the microscopic of the system and our analysis until now is quite general. To proceed to evaluate (\ref{LR}), we need to make some specific reference to the microscopic Hamiltonian and study properties such as the energy spectrum of the quasiparticle. We first notice that all terms in the sum in (\ref{LR}) are negative and so they all contribute to the sum constructively. Let's start by looking into the contribution of the lowest energy eigenstates to the sum which corresponds to bound quasiparticle states. For this, we need to explicitly consider the wave functions and energy spectrum of the bound states. For simplicity, we restrict ourselves to the simplest case with zero superfluid winding number in the absence of any magnetic flux. Generalizations to other cases will not change the qualitative result. \\

We are considering a $2N+1$-particle superfluid system whose condensate forms BCS s-wave Cooper pairs. In the absence of any external potential field, the low-energy eigenstates in the BCS mean-field framework are approximated by the $2N$-particle BCS ground state (particle-number conserving BCS ground state) with a particle-number conserving BdG quasiparticle added:
\begin{eqnarray}
|\mathrm{GS}\rangle_{2N+1}=\alpha^\dagger_{k,\sigma}|\mathrm{GS}\rangle_{2N} \label{GS_odd_uniform}
\end{eqnarray}
with
\begin{eqnarray}
\alpha^\dagger_{k\sigma}=u_ka^\dagger_{k\sigma}+\sigma v_ka_{-k,-\sigma}C^\dagger \label{alpha_BdG}
\end{eqnarray}
and 
\begin{eqnarray}
|\mathrm{GS}\rangle_{2N}=\mathcal{N}C^{\dagger N}|\mathrm{vac}\rangle, C^\dagger=\sum_k c_ka^\dagger_{k\uparrow}a^\dagger_{-k\downarrow}, \label{GS_even}
\end{eqnarray}
where $|\mathrm{vac}\rangle$ denotes the particle vacuum state, $\mathcal{N}$ is the normalization factor, $C^\dagger$ creates Cooper pairs in the BCS ground state, and the coefficients $c_k$ are given by
\begin{eqnarray}
c_k=v_k/u_k, \hspace{5pt} u_k, v_k=\sqrt{1/2}(1\pm\epsilon_k/E_k), \label{coef_BCS}
\end{eqnarray}
where $\epsilon_k=\hbar^2(k^2-k_F^2)/2m$, $E_k=\sqrt{\epsilon_k^2+|\bigtriangleup|^2}$, $k_F$ is Fermi wave vector and $\bigtriangleup$ is the BCS energy gap. A value of $k$ near the Fermi wave vector corresponds to near gap low-energy excitations. \\

Adding a weak Zeeman field $B(z)$ as a function only of $z$ (longitudinal direction along annulus), we get a Zeeman term in the Hamiltonian as
\begin{eqnarray}
\sum_i\sigma_iV(z_i), \hspace{5pt} V(z)=-\mu B(z), \label{Hamiltonian_Zeeman}
\end{eqnarray}
where $\mu$ is the magnetic moment of the particles and $\sigma_i$ is projection of spin along axis of $B$. \\

We consider a regime of the Zeeman field such that a BdG quasiparticle is trapped well within the Zeeman potential and at the same time the condensate can be considered as unaffected. So we require the extension of the Zeeman field $d$ to be much larger than the coherence length $\xi$ and its strength $V_0$ much smaller than $\bigtriangleup$. To trap a quasiparticle within its extension, we require the kinetic energy cost inside the trap be smaller than binding energy, $\hbar v_F/d<V_0$, which is consistently satisfied by the previous requirement of $d\gg \xi$. Finally, the Zeeman field extension should be much smaller than the circumference of the annulus $L=2\pi R$. \\

Let's choose the direction of the Zeeman field to be pointing up, so that the localized quasiparticle has spin up in the lowest two energy eigenstates (doublet). \cite{Leggett_Lin} Due to the Zeeman term, the plane-wave states of the BdG quasiparticle with different momenta will scatter into each other and we expect a bound quasiparticle to be formed out of linear combination of $\alpha^\dagger_{k\uparrow}$ with $k$ around Fermi wave vector $\pm k_F$. Ignoring normal reflection, we can consider wave packet around either $k_F$ or $-k_F$ and the resulting wave functions are described by Andreev bound states \cite{Lin_Leggett}. As bound quasiparticle states with wave vectors in $+z$ and $-z$ directions are degenerate, we can't use them to evaluate the sum in (\ref{LR}) which would result in divergence. So we have to take into full account of the normal reflection which is usually ignored in discussions of Andreev problems. Although the normal reflection splits the lowest two energy levels and avoids divergence, there's something wrong with the energy splitting as described by the BdG equations. For a smooth Zeeman trap varying at length scale much larger than the coherence length, the coupling due to the normal reflection is exponentially small determined by the ratio of Zeeman length scale to the inverse of the Fermi wave number, i.e., it scales as $\mathrm{exp}(-k_Fd)$. On the other hand, we can show that the numerator $|\langle0|J|1\rangle|^2$ remains finite in the limit $d/\xi \rightarrow \infty$, yielding divergent contribution to the sum. This violation of the sum rule (\ref{LR_f}) is closely related to violation of continuity condition and corresponding particle number conservation in the mean-field approach. This can be seen as follows. If the energy splitting of the doublet is exponentially small and suppose the quasiparticle is in the quasi-ground state with wave vector centered around Fermi wave vector $k_F$, then the state is quasi-stationary with lifetime of order the inverse of the exponentially small energy splitting. Now by the continuity condition of particle flow, the divergence of the particle number current should be exponentially small everywhere in the annulus. But this is in contradiction to the corresponding many-body state. As the quasiparticle is localized inside the Zeeman trap, the current outside the trap is zero (remember that the many-body state we are considering has vanishing superfluid velocity) and the current inside is of order $v_F/d$. So the divergence of current is much larger than required by the continuity condition \cite{continuity}. This suggests necessity of going beyond the BdG equations to enforce particle number conservation in order to satisfy the f-sum rule. \\

The reason the BdG approach fails in the above analysis is due to neglecting the superfluid condensate and treating the system as an effective single particle problem associated with the quasiparticle that doesn't preserve particle number conservation. Hence we'll make a variational ansatz to the particle-number conserving many-body ground state in order to recover the continuity condition, though systematic constructions of 'post-BdG' formalism are lacking. To avoid finite current divergence at the edges of the Zeeman trap for the bound quasiparticle approximate energy eigenstate with wave vector around $k_F$, it's natural to imagine some counterflow from the superfluid condensate inside the trap to cancel the current flow induced by the bound quasiparticle. Although neither the condensate nor the quasiparticle satisfies the continuity condition alone, the combination of the two as a whole does. Energetically, the most economical way to generate superfluid flow is to twist the superfluid phase. Hence, it suffices to multiply the 2N-particle ground state wave function (the $2N+1$-particle wave function is obtained by applying the particle-number conserving quasiparticle operator to the $2N$-particle state) by a phase factor $\mathrm{exp}(-i\sum_jf(z_j))$ to yield zero current flow throughout the annulus, where \cite{Lin_Leggett}
\begin{eqnarray}
f(z)=\frac{k_FL}{2N}\int|\Psi_{\mathrm{Sch}}(z)|^2dz, \label{f_z}
\end{eqnarray}
where $L$ is the circumference of the annulus, $\Psi_{\mathrm{Sch}}(z)$ is the solution to a Schrodinger equation derived from the BdG equation for the quasiparticle and is related to the particle and hole component $u(z)$ and $v(z)$ via $u(z)\approx v(z)=\mathrm{exp}(ik_Fz)\Psi_{\mathrm{Sch}}(z)$ (z is the coordinate along the annulus). \\


We see that in the modified state, the superfluid condensate is deformed in the region where the quasiparticle is localized. For the true energy ground state doublet, the quasiparticle wave function is linear combination of two approximate quasiparticle energy eigenstates with wave vectors centered around $\pm k_F$. Once the superfluid condensate deformation is  taken into account, the quasiparticle wave function is entangled with the condensate in the energy eigenstate doublet, a new feature absent in the standard BdG approach. \\

\section{Lower bound on Berry phase at non-zero superfluid velocities} \label{lower bound}

In this section, we estimate a lower bound on the Berry phase change away from integer or half-integer magnetic fluxes, i.e., the sum in equation (\ref{LR}). We will do this by two different approaches and show that they give the same form of the lower bound, one which is independent of the Zeeman trap strength. In the first approach, we apply a Cauchy-Schwartz (CS) inequality. As we notice in evaluating the sum in equations (\ref{LR_1})-(\ref{LR_f}) that we want to avoid evaluating the angular momentum in terms of the azimuthal coordinates, as their definition is subtle in the annular geometry, we need to apply the CS inequality consecutively to obtain the following inequality (Appendix \ref{CS})
\begin{eqnarray}
\sum_n\frac{|\langle0|J|n\rangle|^2}{E_n-E_0} > -\frac{\langle[J,[J,H]]\rangle^3}{\langle[J,H]^2\rangle^2}. \label{CS_lower}
\end{eqnarray}

Evaluating $[J,H]$, we get
\begin{eqnarray}
[J,H]=-iR\int_0^L ds \rho_{\sigma_z}(s)\partial/\partial_sV_z(s)ds, \label{L_H_commut}
\end{eqnarray}
where $R$ is the radius of the annulus , $L=2\pi R$ is its circumference, $V_z(s)$ is the Zeeman field, $\rho_{\sigma_z}(s)$ is spin density in the direction in which the Zeeman field is oriented.  \\

For a wide and weak Zeeman field, we expect a net spin localized in the trap. If we model the bottom of the trap by a harmonic oscillator, $[J,H]$ becomes
\begin{eqnarray}
[J,H]=-iRk\int dr rS_z(r), \label{J_H_commut}
\end{eqnarray}
where we have used $S_z(r)$ to represent the localized spin density (note it is just $|u(r)|^2+|v(r)|^2$ with $u(r)$ and $v(r)$ particle and hole wave functions of the bound quasiparticle, see also \cite{Lin_Leggett}), $k$ is the strength of the oscillator, and $[J,[J,H]]$ is evaluated from (\ref{J_H_commut}) to be
\begin{eqnarray}
[J,[J,H]]=-R^2k\int dr S_z(r)=-R^2k, \label{JJ_H_commut}
\end{eqnarray}
where the normalization condition $\int dr S_z(r)=1$ has been used. \\

The denominator of the rhs of equation (\ref{CS_lower}) can be easily evaluated for a particle in a harmonic oscillator trap to be
\begin{eqnarray}
\langle[J,H]^2\rangle^2=\frac{k^3R^4}{4m^*}, \label{J_H_2}
\end{eqnarray}
where $m^*$ is the effective mass. To obtain the effective mass, we need to explicitly write down the BdG equation for the localized quasiparticle, which will be done shortly in the second approach. For now, we just quote the value $m^*=\frac{\bigtriangleup}{2\epsilon_F}m$ with $m$ the bare particle mass, $\bigtriangleup$ the superconducting gap and $\epsilon_F$ the Fermi energy. \\

Substituting (\ref{J_H_commut}) and (\ref{JJ_H_commut}) into (\ref{CS_lower}) we obtain the lower bound 
\begin{eqnarray}
\sum_n\frac{|\langle0|J|n\rangle|^2}{E_n-E_0} >4R^2m^*. \label{CS_f}
\end{eqnarray}
Taking into account the units (we want to get a dimensionless quantity, and note that the dimension of the above expression is angular momentum$^2$/energy which has units $mR^2$), the rhs of (\ref{CS_f}) becomes $4m^*/m=2\bigtriangleup/\epsilon_F$. Hence, we see that the deviation of the Berry phase away from the integer or half-integer magnetic fluxes has lower bound $-(4\bigtriangleup/\epsilon_F)\delta\Phi$ in the sense that its magnitude is bounded below by $(4\bigtriangleup/\epsilon_F)|\delta\Phi|$ but the deviation has the opposite sign to $\delta\Phi$ (keep in mind the extra factor of two in the sum (\ref{LR})).\\

Now, let's evaluate the lower bound directly from the BdG equation for the bound quasiparticle. Assuming the $2N$-particle ground state wave function is unchanged by the Zeeman trap, we can write down the effective equation obeyed by the bound quasiparticle 
\begin{eqnarray}
E_kC_k-\sum_{k'}V_{k-k'}C_{k'}=EC_k, \label{qp_momentum}
\end{eqnarray}
where $V_{k-k'}$ is the Fourier component of the Zeeman trap and the bound quasiparticle state is given by a linear superposition of plane wave quasiparticle states as
\begin{eqnarray}
\alpha^\dagger=\sum_kC_k\alpha^\dagger_{k\uparrow}. \label{alpha_dagger}
\end{eqnarray}

Expanding the excitation energy $E_k=\sqrt{\epsilon_k^2+\bigtriangleup^2}$ to the lowest order in $\bigtriangleup/\epsilon_F$ and taking into account only wave numbers around $\pm k_F$ as they are dominant in the bound quasiparticle state, we get an effective time-independent Schrodinger equation obeyed by the bound quasiparticle with wave numbers in either direction
\begin{eqnarray}
\frac{1}{2m^*}\frac{d^2}{dz^2}g(z)-V(z)g(z)=Eg(z), \label{TISE}
\end{eqnarray}
where the effective mass is $m^*=(\bigtriangleup/2\epsilon_F)m$, $e^{i\pm k_Fz}g(z)=\sum_{k,\pm k_F}C_k\mathrm{exp}{ikz}$ (the sum over k is near either $k_F$ or $-k_F$). \\

When the superfluid has non-zero winding number around the annulus and is also put in the presence of finite magnetic flux, the effective equation given by (\ref{qp_momentum}) can be generalized by replacing the momentum quantum number by an angular momentum quantum number, and the corresponding excitation energy becomes
\begin{eqnarray}
E_l=\sqrt{\epsilon_l^2+\bigtriangleup^2}+(l_0+2\Phi)(l-\frac{l_0}{2}), \label{E_l}
\end{eqnarray}
where $l_0$ is the condensate winding number, $\Phi$ is the magnetic flux and $\epsilon_l=(l-\frac{l_0}{2})^2-\mu$ with $\mu$ chemical potential.  \\

If we ignore mixing of the wave numbers from around $k_F$ and around $-k_F$, the effective equation is again given by (\ref{TISE}) but in the presence of an effective vector potential. Expanding the energy $E_l$ in terms of $\epsilon_l/\bigtriangleup$, the kinetic terms read
\begin{eqnarray}
E_l&=&\sqrt{\epsilon_l^2+\bigtriangleup^2}+(l_0+2\Phi)(l-\frac{l_0}{2}) \nonumber \\
&\approx&\frac{\epsilon^2_l}{2\bigtriangleup}+(l_0+2\Phi)(l-\frac{l_0}{2})+\bigtriangleup \nonumber \\
&=&\frac{l_{F\pm}^2}{2\bigtriangleup mR^2}(l-l_{F\pm})^2+l(l_0+2\Phi)-l_0(l_0+2\Phi)+\bigtriangleup. \label{kinetic}
\end{eqnarray}
From the above equation, we obtain the average angular momentum (which is equal to the angular momentum which minimizes the kinetic energy)
\begin{eqnarray}
\langle l\rangle= l_{F\pm}-(l_0+2\Phi)\frac{\bigtriangleup}{2\epsilon_F}, \label{l_ave}
\end{eqnarray}
where $l_0+2\Phi$ equals the superfluid velocity, $l_{F+}+l_{F-}=l_0$ ($l_{F\pm}$ refer to positive and negative angular momentum at the Fermi level respectively).\\

The Berry phase can be straightforwardly shown (by an argument similar to that for the general case leading to equation (\ref{Berry}), i.e., the bound quasiparticle wave function depends on the Zeeman trap location parametrized by $\theta_0$, so the Berry phase is related to its angular momentum) to be equal to the average angular momentum of the bound quasiparticle
\begin{eqnarray}
\phi/2\pi=\langle l\rangle. \label{phi_qp}
\end{eqnarray}

Ignoring the redistribution of the magnitudes of $C_l$ around $l_{F+}$ relative to those around $l_{F-}$ due to the magnetic flux change $\delta\Phi$, we get by combining equation (\ref{phi_qp}) and equation (\ref{l_ave}) a lower bound for the Berry phase change due to $\delta\Phi=\frac{1}{2}+\Phi$ (coming from the term linear in $\Phi$ in equation (\ref{l_ave}))
\begin{eqnarray}
|\delta\phi/2\pi|>(\bigtriangleup/\epsilon_F)|\delta\Phi| \label{delta_phi}
\end{eqnarray}
where again $\delta\phi$ has the opposite sign to $\delta\Phi$ as discussed below equation (\ref{CS_f}).\\

Hence, we see that the two approaches give a similar lower bound and the former approach gives a stronger bound (four times that of the latter bound). \\

In the latter approach, we further see that the redistribution of the $C_l$ around positive and negative Fermi momentum due to finite superfluid velocity (i.e. finite $l_0+2\Phi$) is dependent on the potential energy saving by scattering between opposite Fermi momenta through the Zeeman trap (without the trap, all coefficients will be around either positive or negative Fermi momentum depending on the direction of the superfluid velocity). So for the many-body wave functions constructed by the BdG solutions, the Berry phase at general magnetic fluxes is non-universal and depends on parameters of the system such as the Zeeman trap strength. \\

In this section, we have shown that the modification of the Berry phase at non-zero superfluid velocities is non-zero. Therefore the Berry phase is affected by the motion of the superfluid, the effect is bounded from below by an amount dependent on quantities such as the superfluid gap and the Fermi energy. Furthermore, the Berry phase is non-universal and is dependent on both the superfluid properties such as the superfluid gap and the Zeeman trap. We shall now verify the non-universality of the Berry phase by focusing on a specific form of the Zeeman potential. \\

\section{Berry phase evaluated for a square-well Zeeman potential} \label{square}

In this section, we evaluate the Berry phase quantitatively for a specific shape of Zeeman trap, a square-well trap. (Note that the considerations of ref.[6] concerning the exponentially small splitting of the lowest two states do not apply to this case). We develop a different approach for calculating the Berry phase based on a work-energy relationship. The advantage of this approach is that the Berry phase can be evaluated from the energy of the bound quasiparticle which is properly addressed by the BdG equations, without the need to explicitly refer to quantities more difficult to calculate such as the angular momentum and many-body wave functions. \\

We have shown from symmetry that the Berry phase is $\pi$ (equal to the AB phase) at magnetic flux $\Phi=-\frac{1}{2}$ (in units of $h/|e|$) and vanishing superfluid velocity. We are interested in knowing the evolution of the Berry phase as we change the magnetic flux while keeping the winding number $l_0$ fixed. As we change the magnetic flux, a voltage is generated around the annulus. The amount of work done to the system is determined by the voltage and the current. Since work is equal to energy change, we can establish a relationship between the Berry phase (in terms of current) and variation of the quasiparticle energy. \\

The work-energy equations for the system with $2N$ and $2N+1$ particles are
\begin{eqnarray}
\int_{\Phi_i}^{\Phi_f}(L_{2N}+2N\Phi)d\Phi&=&E_{2N}(\Phi_f)-E_{2N}(\Phi_i) \nonumber \\
\int_{\Phi_i}^{\Phi_f}(L_{2N+1}+(2N+1)\Phi)d\Phi&=&E_{2N+1}(\Phi_f)-E_{2N+1}(\Phi_i), \label{work-energy}
\end{eqnarray}
where $L$ denotes total angular momentum, $E$ refers to total energy of the system and $\Phi$ is external magnetic flux. All three quantities are dimensionless in units $\hbar$, $\hbar^2/2mR^2$ ($R$ - annulus radius) and $h/|e|$, respectively. Subtracting the two equations, we obtain
\begin{eqnarray}
\int_{\Phi_i}^{\Phi_f}(L_{2N+1}-L_{2N}+\Phi)d\Phi=(E_{2N+1}(\Phi_f)-E_{2N}(\Phi_f))-(E_{2N+1}(\Phi_i)-E_{2N+1}(\Phi_i)). 
\label{work-energy_qp}
\end{eqnarray}

Differentiating equation (\ref{work-energy_qp}), we get
\begin{eqnarray}
L_{2N+1}-L_{2N}=-\Phi+dE(\Phi)/d\Phi, \label{work-energy_qp_f}
\end{eqnarray}
where $E(\Phi)\equiv E_{2N+1}(\Phi)-E_{2N}(\Phi)$ is the quasiparticle energy. \\

Since the Berry phase $\phi$ is equal to $2\pi L$, we get the Berry phase from (\ref{work-energy_qp_f})
\begin{eqnarray}
\phi/2\pi=-\Phi+dE(\Phi)/d\Phi. \label{Berry_energy}
\end{eqnarray}
The first term on rhs of equation (\ref{Berry_energy}) is just the AB phase, the second term is the correction due to the energy dependence on the magnetic flux. \\

We now solve the BdG equations to find $E(\Phi)$. In order to find a simple analytic expression, we choose a square shaped Zeeman trap, i.e. it is constant between $0$ and $\theta=\theta_L$ ($\theta$ is the azimuthal coordinate parametrizing the annulus) and zero elsewhere. We also ignore any spatial inhomogeneity of the gap magnitude, a valid approximation for weak enough Zeeman field. \\

We consider a magnetic flux around $-1/2$, where the BdG equation reads
\begin{eqnarray}
((l-\frac{1}{2})^2-\mu)u+\bigtriangleup e^{i\theta}v&=&(E+V)u \nonumber \\
-((l+\frac{1}{2})^2-\mu)v+\bigtriangleup e^{-i\theta}u&=&(E+V)v, \label{BdG_1/2_B}
\end{eqnarray}
where $V$ is the Zeeman potential energy which is zero between $\theta_L$ and $2\pi$ and equal to $V$ elsewhere. $\bigtriangleup$ is constant to a good approximation and it can be made to be real, $l$ is the angular momentum quantum number, $\mu$ is chemical potential. Let's make the gauge transformation $u=\tilde{u}e^{i\theta/2}$, $v=\tilde{v}e^{-i\theta/2}$. After the transformation and renaming $\tilde{u}$, $\tilde{v}$ to $u$, $v$, we get
\begin{eqnarray}
(l^2-\mu)u+\bigtriangleup v&=&(E+V)u \nonumber \\
-(l^2-\mu)v+\bigtriangleup u&=&(E+V)v. \label{BdG_1/2_B_transf}
\end{eqnarray}
Now $u$ and $v$ become anti-periodic around the annulus, which doesn't affect a bound solution as it is localized. \\

When the magnetic flux is $\Phi=-\frac{1}{2}+\lambda$, the BdG equation becomes
\begin{eqnarray}
((l+\lambda)^2-\mu)u+\bigtriangleup v&=&(E+V)u \nonumber \\
-((l-\lambda)^2-\mu)v+\bigtriangleup u&=&(E+V)v, \label{BdG_0_transf}
\end{eqnarray}
where again $u$ and $v$ satisfy an anti-periodic boundary condition. The BdG equation (\ref{BdG_0_transf}) has the following symmetry: $\lambda\rightarrow-\lambda$, $E\rightarrow E$ for $(u,v)\rightarrow(u^*,v^*)$. Thus, the lowest energy eigenvalue $E_0(\lambda)$ is symmetric around $\lambda=0$. So $dE(\Phi)/d\Phi=0$ at $\Phi=-1/2$, and equation (\ref{Berry_energy}) gives just an AB phase at $\Phi=-1/2$, consistent with our previous general considerations based on symmetry (cf. Section \ref{Integer Flux}).\\

If we ignore mixing between positive and negative momenta, it's relatively straightforward to show that the lowest energy doublet is degenerate at $\Phi=-1/2$ and the energies are linear in $\lambda$ with slope $\pm2l_F$ ($l_F^2=\mu$) (see appendix \ref{square well} for a detailed derivation). When mixing is taken into account, an energy gap is opened up at the crossing point and we can write down the energy spectrum for small $\lambda$ as
\begin{eqnarray}
E(\lambda)=\pm\sqrt{(2l_F\lambda)^2+\eta^2}, \label{doublet_splitting}
\end{eqnarray}
where $\eta$ is the energy gap at $\Phi=-1/2$. So the derivative of the lowest energy with respect to $\lambda$ is 
\begin{eqnarray}
dE/d\lambda=-4l_F^2\lambda/\sqrt{(2l_F\lambda)^2+\eta^2}. \label{dEdlambda}
\end{eqnarray}

Equation (\ref{dEdlambda}) shows explicitly that the Berry phase is non-universal, depending on both the Fermi energy and the energy gap of the doublet splitting which is determined by both the Zeeman trap and the superfluid gap etc. It is further interesting to note the follow points concerning equation (\ref{dEdlambda}). Firstly, it satisfies the f-sum rule and its upper bound is just half of the upper bound on the f-sum rule for a true 1D system since the magnitude of the rhs is bounded by $2l_F$ which is equal to the total number of particles for each spin. The physical system of interest is only quasi-1D and therefore the upper bound is smaller than half the upper bound on the f-sum rule. Secondly, comparing (\ref{dEdlambda}) with (\ref{LR}), we see that in the linear regime where $2l_F\lambda\ll\eta$, the Berry phase for a square well takes the same form as the doublet contribution to the sum in the linear response formula (up to a factor of 2) at the BdG level (i.e., many-body deformation is not taken into account). \\

Finally, we note that although in calculating the Berry phase, we have used the BdG equations to evaluate the quasiparticle energy spectrum, the formula (\ref{Berry_energy}) for the Berry phase is derived in a way that fully respects particle number conservation and furthermore relies only on the system energy spectrum. This explains why the expression (\ref{dEdlambda}), obtained from the BdG equations, satisfies the f-sum rule since the BdG equations can address the system energy levels properly. This is in contrast to using the expression (\ref{LR}) to evaluate the Berry phase, where evaluation of the numerators relies on the system eigenstate wave functions. As demonstrated in section \ref{continuity condition}, the bound quasiparticle state doublet violates the continuity condition in the BdG framework, resulting in violation of the f-sum in evaluating the Berry phase. \\

\section{Summary and Conclusions} \label{summary}

The Berry phase associated with transport of a quasiparticle around a superfluid vortex is surprisingly a rather subtle problem, due to the effect of the superfluid condensate when it is moving relative to the bound quasiparticle. At integer or half-integer magnetic fluxes (in units of $h/|e|$), a general argument STET the exact result that the Berry phase is equal to the AB phase, a result independent of any approximation (Section \ref{Integer Flux}). Away from the integer or half-integer fluxes, in order to take into account the condensate effect, it is crucial to respect particle number conservation. This is manifest throughout our discussions with a variety of approaches. \\

If we regard the system ground state as a single-particle problem and map it into an effective spin-1/2 in a magnetic field, we explicitly break particle number conservation since the spin degree of freedom corresponds to particle and hole components of the quasiparticle. The Berry phase is found to be equal to $\pi$, irrespective of the external magnetic flux through the annulus (Section \ref{BdG}). In a different approach, the Berry phase can be related to the total angular momentum of the system in the one-dimensional limit (annulus thickness much smaller than its radius) and making use of the continuity condition, the Berry phase can be obtained by evaluating the commutation relation between the localized quasiparticle operator and the local current operator (Section \ref{continuity condition}). In the standard particle-number non-conserving formalism, the commutator vanishes as the Cooper pair operator associated with hole part of the quasiparticle is neglected. Therefore, the Berry phase is found to be equal to AB phase at any magnetic flux. This conclusion is in contradiction with that obtained from the effective spin model away from integer or half-integer magnetic fluxes. The inconsistency between the two approaches come from breaking particle number conservation, resulting in an incorrect Berry phase away from integer or half-integer magnetic fluxes by both approaches.\\

The effect of particle number conservation is particularly manifest in linear response theory (or alternatively, from adiabatic perturbation theory in a rotating frame in which the Zeeman trap is at rest) in which the Berry phase is written in the sum given by equation (\ref{LR}). This sum puts a upper bound on the Berry phase equal to the total number of particles of the system, i.e., $2N+1$. Examination of the first term in the sum shows that it violates the upper bound, as the energy splitting between the lowest two energy levels is exponentially small for a wide smooth Zeeman trap. This violation is closely related to the violation of the continuity condition for approximate eigenstates made out of a linear combination of the doublet eigenstates, i.e, for approximate states traveling in either direction inside the trap. To resolve this issue, we propose to modify the condensate wave function in response to quasiparticles traveling in either directions such that the quasiparticle wave function and the condensate wave function is entangled (Section \ref{continuity condition}). The modified many-body state respects continuity condition and removes the divergence in the sum. This modification goes beyond the BdG framework in which the quasiparticle wave function is not entangled with the condensate wave function. This can have important implications for Majorana zero modes, as relevant many-body ground states may be subject to modifications beyond the BdG description, potentially affecting the properties of Majorana zero modes as given by the BdG description. \\

We have also estimated a lower bound on the Berry phase modification away from the integer or half-integer magnetic fluxes in order to establish the non-zero modification of the Berry phase due to the superfluid flow. From both the CS inequality and directly evaluating the quasiparticle wave function as a solution to the effective time-independent Schrodinger equation, we get a lower bound of order $\bigtriangleup/\epsilon_F$. This lower bound rules out the Berry phase of $\pi$ given by the effective spin model. Furthermore, in evaluating the quasiparticle wave function, we see that the Berry phase is non-universal depending on the parameters of the system and the Zeeman trap. Of course, this conclusion is based on a many-body wave function constructed from the BdG solutions and its validity beyond the BdG approximation is not completely clear. \\

Finally, we develop a formula for the Berry phase which depends only on the energy spectrum of the system, thus independent of particle number conservation. We use this formula to explicitly calculate the Berry phase for a square-well Zeeman trap and obtain a non-universal result similar to the contribution of doublet to the sum in (\ref{LR}) within the BdG formalism. It's tempting to believe that the non-universality of the Berry phase at general magnetic flux is valid even beyond the BdG approximation since we don't expect the energy spectrum to modify significantly beyond the BdG equations. \\

In this paper, we hope to have convinced the reader that careful application of BdG equations is needed to evaluating the Berry phase associated with the transport of a quasiparticle around a superfluid vortex, and that formalism beyond the BdG framework may be necessary for obtaining a physically correct result. Particle number conservation plays an important role in the Berry phase when the superfluid condensate is moving. The lesson we learn from this study may shed important light on understanding the properties of Majorana zero modes beyond the BdG framework. \\

\section*{Acknowledgement}
This work was supported by the National Science Foundation through grant NSF-DMR-09-06921.

 \appendix
 \appendixpage
\begin{appendices}
 \section{Derivation of lower bound on Berry phase with Cauchy-Schwartz inequality} \label{CS}
 
 In this appendix, we derive equation (\ref{CS_lower}) in the main text by applying the Cauchy-Schwartz (CS) inequality consecutively. For convenience, we define the following moments
\begin{eqnarray}
I_k\equiv \sum_n|\langle0|J|n\rangle|^2(E_n-E_0)^k \label{mom_k}
\end{eqnarray}
The CS inequality then reads
\begin{eqnarray}
I_k^2\leq I_{k-1}I_{k+1} \label{CS_ineq}
\end{eqnarray}

Applying (\ref{CS_ineq}) consecutively to evaluating the lower bound to the sum $\sum_n|\langle0|J|n\rangle|^2/(E_n-E_0)$, we get
\begin{eqnarray} 
\sum_n\frac{|\langle0|J|n\rangle|^2}{E_n-E_0}&=&I_{-1} \geq\frac{I^2_0}{I_1} =\frac{I^2_0I_2}{I_1I_2}\geq\frac{I_0I_1^2}{I_1I_2} =\frac{I_0I_1I_2}{I_2^2} \geq\frac{I_1^3}{I_2^2}  \label{CS_ineq_f}
\end{eqnarray}
Writing the rhs of the above inequality in terms of commutators, we arrive at equation (\ref{CS_lower}). End of Derivation.\\

 \section{Energy spectrum of bound quasiparticle in square-well Zeeman potential} \label{square well}

In this appendix, we derive the energy spectrum of a localized quasiparticle in a square-well Zeeman potential trap. The derivation is given by Appendix A in \cite{Lin_Leggett} for the case without any external magnetic flux and at null superfluid winding number. For the reader's convenience, we include it here. The generalization to the case of interest where the superfluid velocity is non-zero is straightforward and we will give the corresponding expression directly. \\

For simplicity of notation, we'll present details at magnetic flux $\Phi=-1/2$, i.e., we'll solve equation (\ref{BdG_1/2_B_transf}). The corresponding result for general magnetic flux $\Phi=-1/2+\lambda$ (corresponding to equation (\ref{BdG_0_transf})) will be listed without giving an explicit derivation. \\

We solve equation (\ref{BdG_1/2_B_transf}) by expanding the wave vector around $p_F$ (we consider linear momentum here, which is equivalent to angular momentum in 1D). Inside the well, i.e., between $\theta=0$ and $\theta=\theta_L$, the BdG quasiparticle states are plane waves; outside the well, they decay exponentially. We solve them by matching the boundary conditions at the edge of the well. It is convenient to write the ratio of $u$ and $v$ as
\begin{eqnarray}
\frac{v}{u} &=& \frac{\bigtriangleup}{E+\Omega_o^{\pm}}=F_o^{\pm} \nonumber \\
\frac{v}{u}&=&\frac{\bigtriangleup}{E+V+\Omega_i^{\pm}}=F_i^{\pm}, \label{v_u}
\end{eqnarray}
where subscripts $o$ and $i$ refer to outside and inside the Zeeman trap respectively, we have omitted subscripts $o$, $i$ and superscripts $\pm$ for $u$ and $v$ for notational simplicity.  $\Omega_o^{\pm}=\pm\sqrt{\bigtriangleup^2-E^2}i$, $\Omega_i^{\pm}=\pm\sqrt{(E+V)^2-\bigtriangleup^2}$. Note that $\Omega_o^{\pm}$ is pure imaginary and this is the factor which gives rise to the exponential decay of the bound solutions outside the trap. Strictly speaking, there are eight boundary conditions, four of them coming from the continuity conditions for u and v at the two trap edges, the other half coming from the continuity conditions for the first derivatives of u and v. Since upon Andreev reflection, the momentum change is of order $\bigtriangleup/E_F$ compared to the Fermi momentum, the momenta of the particle and hole plane wave solutions differ only by order $\bigtriangleup/E_F$, relative to the Fermi momentum. If we ignore this small difference, the plane wave solutions for wave vectors in opposite directions become separate.  We only need to match the continuity conditions for u and v (with wave vectors in one direction) and the continuity conditions for the first derivatives are automatically satisfied (this point is discussed in \cite{DG}). Within this approximation, the solutions with momentum in two opposite directions are degenerate, as is the case for the general gradually varying potential discussed in the main text. Consider a solution inside the trap of the form
\begin{eqnarray}
u_i&=&u_i^+e^{ip_i^+\theta}+u_i^-e^{ip_i^-\theta} \nonumber \\
v_i&=&F_i^+u_i^+e^{ip_i^+\theta}+F_i^-u_i^-e^{ip_i^-\theta}, \label{u_v_i}
\end{eqnarray}
where $p_i^\pm=p_F+\frac{\Omega_i^\pm}{v_F}$; in getting the second equation, equation (\ref{v_u}) is used. \\ 

Similarly, the solution outside the trap is given by 
\begin{eqnarray}
u_o&=&u_o^+e^{ip_o^+(\theta-\theta_L)}+u_o^-e^{ip_o^-(2\pi-\theta)} \nonumber \\
v_o&=&F_o^+u_o^+e^{ip_o^+(\theta-\theta_L)}+F_o^-u_o^-e^{ip_o^-(2\pi-\theta)}, \label{u_v_i}
\end{eqnarray}
where $p_o^\pm=p_F+\frac{\Omega_o^\pm}{v_F}$.\\

Now by matching the boundary conditions at $\theta=2\pi$ and $\theta=\theta_L$, we get the following equations at $\theta=2\pi$
\begin{eqnarray}
u_o^+e^{ip_o^+(2\pi-\theta_L)}+u_o^-&=&u_i^+ +u_i^- \nonumber \\
F_o^+u_o^+e^{ip_o^+(2\pi-\theta_L)}+F_o^-u_o^-&=&F_i^+u_i^+ +F_i^-u_i^-. \label{BC_2pi}
\end{eqnarray}
Similarly, the equations at $\theta=\theta_L$ are
\begin{eqnarray}
u_o^+ +u_o^-e^{ip_o^-(\theta_L-2\pi)}&=&u_i^+e^{ip_i^+\theta_L} +u_i^-e^{ip_i^-\theta_L} \nonumber \\
F_o^+u_o^+ +F_o^-u_o^-e^{ip_o^-(\theta_L-2\pi)}&=&F_i^+u_i^+e^{ip_i^+\theta_L} +F_i^-u_i^-e^{ip_i^-\theta_L}. \label{BC_thetaL}
\end{eqnarray}
Neglecting exponentially small terms in equation (\ref{BC_2pi}) and (\ref{BC_thetaL}), we get the following equation
\begin{eqnarray}
\frac{F_i^- -F_o^-}{F_o^- -F_i^+}=\frac{F_i^- -F_o^+}{F_o^+ -F_i^+}e^{i(p_i^--p_i^+)\theta_L}. \label{E}
\end{eqnarray}

This equation can be written as 
\begin{eqnarray}
\mathrm{tan}^{-1}(\frac{\sqrt{\bigtriangleup^2-E^2}}{V-\sqrt{(E+V)^2-\bigtriangleup^2}})-\mathrm{tan}^{-1}(\frac{\sqrt{\bigtriangleup^2-E^2}}{V+\sqrt{(E+V)^2-\bigtriangleup^2}})=-\frac{\sqrt{(E+V)^2-\bigtriangleup^2}}{v_F}\theta_L+n\pi, \nonumber \\
 \label{E_1}
\end{eqnarray}
where $n$ is an integer. \\

We can find simple solutions to (\ref{E_1}) in the limit of a wide trap, i.e., satisfying the condition
\begin{eqnarray}
(E+V)^2-\bigtriangleup^2\ll V^2. \label{wide_trap}
\end{eqnarray}
This condition is equivalent to
\begin{eqnarray}
E+V-\bigtriangleup\sim\frac{v_F^2}{\bigtriangleup\theta_L^2}\ll \frac{V^2}{\bigtriangleup} \nonumber \\
\therefore \theta_L\gg \theta_V, \label{wide_trap_1}
\end{eqnarray}
where $\theta_V=v_F/V$ is the length scale associated with the trap strength ($\hbar=1$). \\

Under this condition, we can expand equation (\ref{E_1}) to first order in $\sqrt{((E+V)^2-\bigtriangleup^2)}/V$ and we get
\begin{eqnarray}
\frac{1}{1+\frac{\bigtriangleup^2-E^2}{V^2}}=-\frac{\theta_LV}{2v_F}+\frac{n\pi}{2\sqrt{\frac{(E+V)^2-\bigtriangleup^2}{V^2}}}. \label{E_2}
\end{eqnarray}
Now, consider low bound states such that $(E+V-\bigtriangleup)/V\ll1$ and  $(\bigtriangleup-E)/V\sim 1$. Since the lhs of equation (\ref{E_2}) is much smaller than 1 and the absolute values of both terms on the rhs of equation (\ref{E_2}) are much greater than 1, we could set the lhs to zero. Hence, we arrive at the solution
\begin{eqnarray}
E=\sqrt{(\frac{n\pi v_F}{\theta_L})^2+\bigtriangleup^2}-V \label{solution_E}
\end{eqnarray}

This solution is consistent with the intuitive argument. For low bound states, all the Zeeman energy $V$ is saved since low bound wave functions are completely localized inside the trap, i.e., their range outside the trap is negligible, hence the term $-V$ in equation (\ref{solution_E}). The term $(2n\pi v_F/\theta_L)^2$ in the square root of the above equation is simply the kinetic energy of the quasiparticle since its momentum $\delta p=p-p_F$ is quantized by the trap as $2n\pi/(2\theta_L)$, with $\theta_L$ as the trap width. \\

Our result (\ref{solution_E}) is consistent with the standard picture that due to the Andreev reflection, the quantization length is twice the trap width \cite{Kulik}. \\

The energy spectrum for general magnetic flux $\Phi=-1/2+\lambda$ can be similarly found to be
\begin{eqnarray}
(E+V)^2-4(E+V)p_F\lambda+4p_F^2\lambda^2=(\frac{2\pi np_F}{\theta_L})^2+\bigtriangleup^2 \label{E_flux}
\end{eqnarray}

Taking derivative with respect to $\lambda$, we obtain the result in the main text (cf. equation (\ref{doublet_splitting}) in the absence of energy splitting), namely, near $\lambda=0$, $dE/d\lambda=2p_F$ (here $p_F$ refers to the angular momentum at the Fermi energy with $\hbar=1$, the energy unit is $\hbar^2/2mR^2$ with $R$ radius of annulus).

\end{appendices}


\end{document}